\documentclass[sigconf]{acmart}

\usepackage{booktabs}
\usepackage{algorithm}
\usepackage{algpseudocode}
\usepackage{bbold}
\usepackage{pgfplots}
\usepackage{graphicx}
\usepackage{subcaption}
\usepackage{calrsfs}
\usepackage{amsmath}

\DeclareMathAlphabet{\pazocal}{OMS}{zplm}{m}{n}

\newtheoremstyle{mydef}% style name
{2ex}% above space
{2ex}% below space
{\itshape}% body font
{}% indent amount
{\scshape}% head font
{: }% post head punctuation
{0.5em}% space after theorem head
{}% head spec 
\theoremstyle{mydef}

\usepackage{enumitem}
\setlist[itemize,1]{leftmargin=0.4cm}

\begin{document}

\copyrightyear{2019} 
\acmYear{2019} 
\setcopyright{acmcopyright}
\acmConference[SIGIR '19]{Proceedings of the 42nd International ACM SIGIR Conference on Research and Development in Information Retrieval}{July 21--25, 2019}{Paris, France}
\acmBooktitle{Proceedings of the 42nd International ACM SIGIR Conference on Research and Development in Information Retrieval (SIGIR '19), July 21--25, 2019, Paris, France}
\acmPrice{15.00}
\acmDOI{10.1145/3331184.3331333}
\acmISBN{978-1-4503-6172-9/19/07}

\fancyhead{}

\title{Table2Vec: Neural Word and Entity Embeddings for Table Population and Retrieval}
\fancyhead{}

\author{Li Deng}
\affiliation{%
  \institution{University of Stavanger}
}
\email{ninalx1991@gmail.com}
%\email{l.deng@stud.uis.no}

\author{Shuo Zhang}
\affiliation{%
  \institution{University of Stavanger}
}
\email{shuo.zhang@uis.no}

\author{Krisztian Balog}
\affiliation{%
  \institution{University of Stavanger}
}
\email{krisztian.balog@uis.no}

\begin{abstract}
Tables contain valuable knowledge in a structured form.  We employ neural language modeling approaches to embed tabular data into vector spaces.  Specifically, we consider different table elements, such caption, column headings, and cells, for training word and entity embeddings.  These embeddings are then utilized in three particular table-related tasks, row population, column population, and table retrieval, by incorporating them into existing retrieval models as additional semantic similarity signals.  Evaluation results show that table embeddings can significantly improve upon the performance of state-of-the-art baselines.
\end{abstract}



% TODO: check SIGIR list of topics!
% Topics:	entity-oriented search, information access, intelligent personal assistants, ranking algorithms, semantic search, zero-query and implicit search

\maketitle

\vspace*{-0.75\baselineskip}
\section{Introduction}
\label{sec:int}

Tables contain a vast amount of useful information in the form of structured data. Recently, a growing body of work has developed around leveraging tabular data in various applications~\citep{Zhang:2018:SES, Zhang:2018:AHT, Cafarella:2009:DIR, Pimplikar:2012:ATQ, Yakout:2012:NTE,Zhang:2017:ESA,DasSarma:2012:FRT,Ahmadov:2015:NTE,Sekhavat:2014:KBE, Zhang:2018:OTG}.  In this paper, we focus on three particular table-related tasks: row population, column population, and table retrieval.  All three tasks are performed on \emph{relational tables}, which describe a set of entities placed in a \emph{core column}, along with their attributes in additional columns. 
\emph{Table population} is the task of populating a given seed table with additional elements~\citep{Yakout:2012:NTE,Zhang:2017:ESA}.  Specifically, we address the  \emph{row population} and \emph{column population} tasks proposed in~\citep{Zhang:2017:ESA}.  The former aims to complement the core column of a relational table with additional entities, while the latter aims to complement the header row with additional headings. 
\emph{Table retrieval} is the task of returning a ranked list of tables for a keyword query~\citep{Zhang:2018:AHT}. 

\begin{figure}[t]
    \begin{subfigure}[b]{0.22\textwidth}
        \centering
        \includegraphics[width = \textwidth]{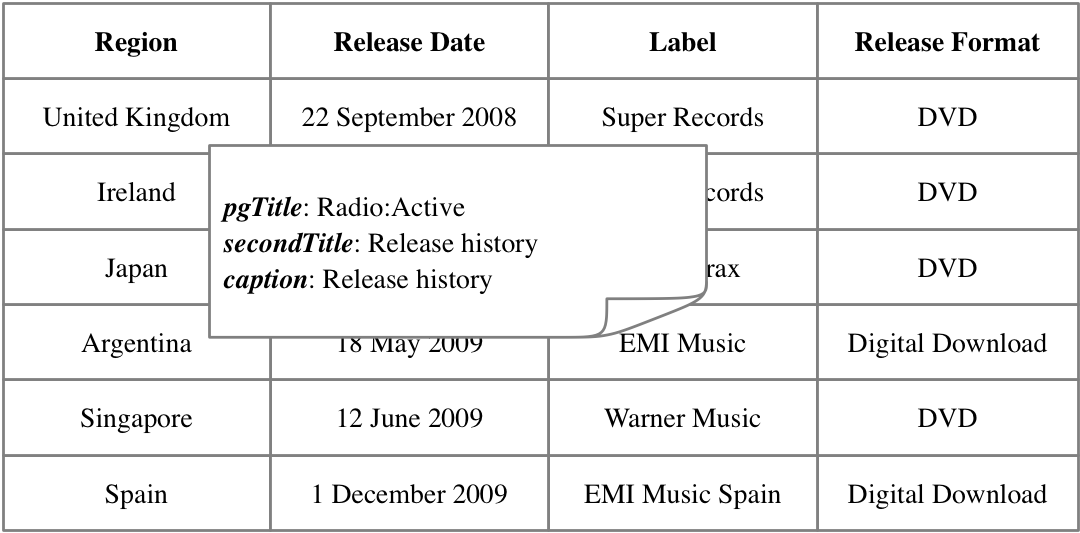}
        \subcaption{\emph{Table2VecW}}
        \label{fig:a}
    \end{subfigure}
    \hspace{0.5em}
    \begin{subfigure}[b]{0.22\textwidth}
        \centering
        \includegraphics[width = \textwidth]{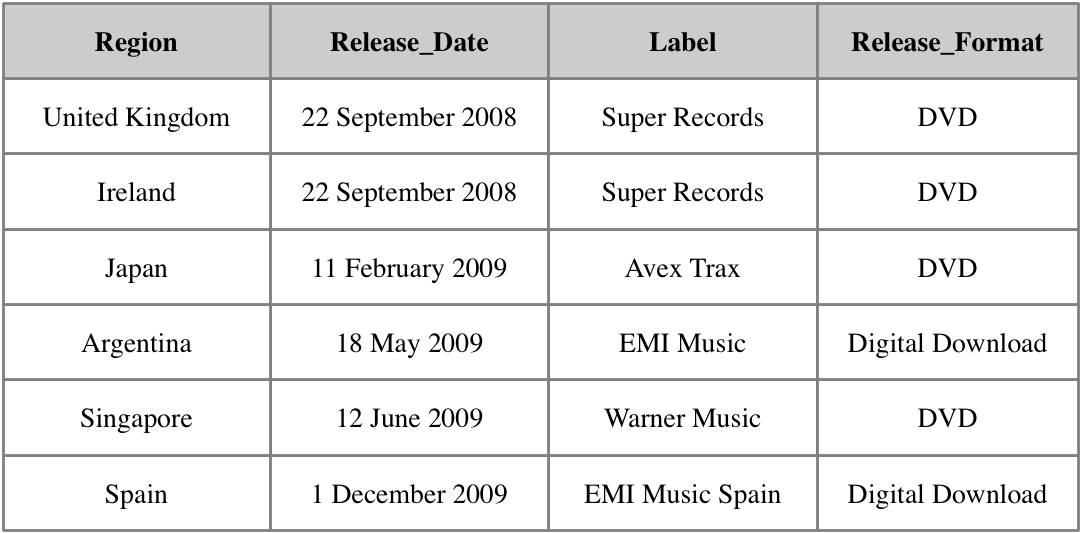}
        \subcaption{\emph{Table2VecH}}
        \label{fig:b}        
    \end{subfigure}
	\\
    \begin{subfigure}[b]{0.22\textwidth}
        \centering
        \includegraphics[width = \textwidth]{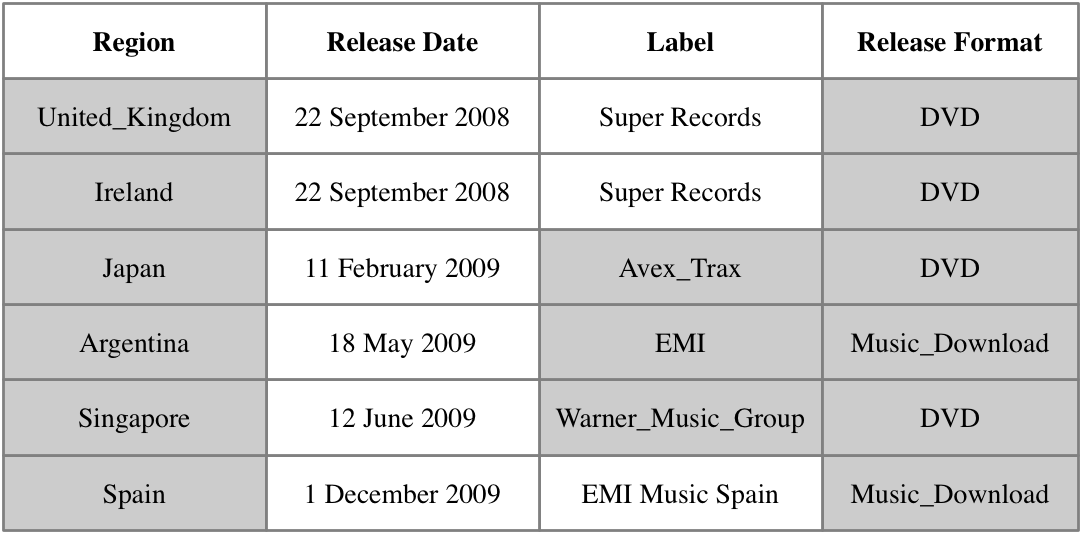}
        \subcaption{\emph{Table2VecE}}
        \label{fig:c} 
    \end{subfigure}
    \hspace{0.5em}
    \begin{subfigure}[b]{0.22\textwidth}
        \centering
        \includegraphics[width = \textwidth]{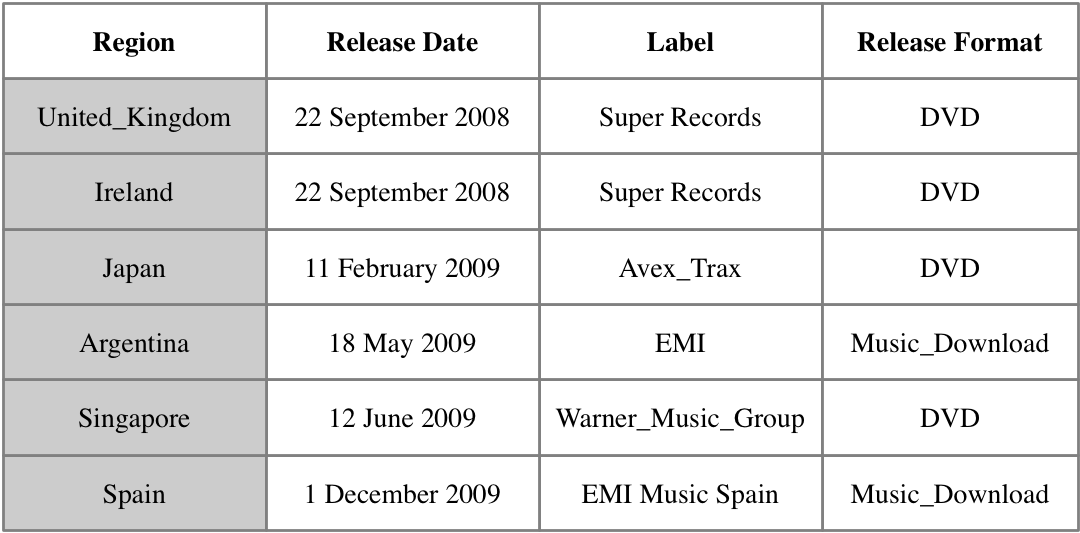}
        \subcaption{\emph{Table2VecE*}}
        \label{fig:d} 
    \end{subfigure}
    \caption{Illustration of different Table2Vec embeddings.}
    \label{fig:table2vec}
    \vspace*{-1\baselineskip}
\end{figure} 

Prior table-related work has considered embeddings, both pre-trained ones and task-specific ones.
For example, \citet{Zhang:2018:AHT} use pre-trained word and entity embeddings for table retrieval. 
\citet{Majid:2018:TTV} develop table embeddings for table classification and \citet{Anna:2017:EMW} train table embeddings for web table entity matching. 
However, to the best of our knowledge, no studies have been conducted on training table embeddings specifically for table population and retrieval tasks.
To fill the gap, we propose Table2Vec, a neural language modeling approach to map different table elements into semantic vector spaces, with specific table-oriented tasks in mind.

In this study, we train four variants of table embeddings by utilizing different table elements. Specifically, word embeddings (Table2VecW) consider all the terms within a table, and are leveraged for table retrieval.  The method employing Table2VecW outperforms a start-of-the-art baseline by over 10\% in terms of NDCG@10.  Interestingly, this is on par with using pre-trained Word2Vec embeddings using Google News data. 
Two different entity embeddings are obtained by considering only core column entities (Table2VecE*) and all table entities (Table2VecE).  Table2VecE* is employed for the row population task.  We show that it significantly outperforms all baselines.  Combining with an effective baseline can lead to further improvements. Table2VecE is employed in table retrieval and can yield minor improvements, albeit those are not statistically significant.
Heading embeddings (Table2VecH) are generated for the column population task by utilizing table headings.  Table2VecH results in substantial and significant improvements over the baseline.  Especially, when the number of seed headings becomes larger, it achieves 40\% relative improvement in NDCG@10 over the baseline.

% \noindent
The resources developed in this work are made publicly available at \url{https://github.com/iai-group/sigir2019-table2vec}.

\vspace*{-0.75\baselineskip}
\section{Training Table2Vec Embeddings}
\label{sec:nwe}

In this section, we first introduce the neural model for training embeddings (Sect.~\ref{sub:nwe:nmt}), and then detail four variants of table embeddings (Sect.~\ref{sub:nwe:wer}).

%How to leverage the table structure and extract the rightful data for constructing neural language models are problems themselves. In this section, we propose four different table raw representations for corresponding Table2Vec embeddings, and use these vector representations for subsequent table population and retrieval tasks. 
%
%\input{other-inputs/fig-emb-exp.tex}

\vspace*{-0.75\baselineskip}
\subsection{Neural Model for Training Embeddings}
\label{sub:nwe:nmt}

We base the training of our embeddings on the skip-gram neural network model of \emph{Word2Vec}~\citep{Mikolov:2013:DRW}. It is a computationally efficient two-layer neural language model that learns the meaning of terms from raw sequences and maps those terms to a vector space, such that similar terms close to each other. 

%Figure~\ref{fig:skip} shows the architecture of the skip-gram model. 
%\input{other-inputs/fig-skip-gram.tex}

More formally, given a sequence of training terms $t_{1}, t_{2}, \ldots ,t_{n}$, the objective is to maximize the average log probability:
\begin{equation}
	\frac{1}{n}\sum_{i=1}^{n}\sum_{-c\leq j\leq c, j \neq 0}\log p(t_{i+j}\vert t_{i}) ~,
\end{equation}
where $c$ is the size of training context, and the probability $p(t_{i+j} \vert t_{i})$ is calculated using the following softmax function:
\begin{equation}
p(t_{o}\vert t_{i})=\frac{{\exp(\vec{v}'_{t_{o}}}^\top \vec{v}_{t_{i}})}{\sum_{t=1}^{V}{\exp(\vec{v}'_{t}}^\top \vec{v}_{t_{i}})} ~,
\end{equation}
where $V$ is the size of vocabulary, and $\vec{v}_{t_i}$ and $\vec{v}'_{t_o}$ are the input and output vector representations of term $t$, respectively.  %Note that there are actually two representations of a term (apart from the one-hot vector): $\vec{v}_{t}$ is the embedded vector for $t$ as the center word and $\vec{v}'_{t}$ is the vector representation for $t$ as the context word.% 
Semantically similar terms share more similar vector representations; accordingly, the dot product between those vectors results in higher values, which means higher probabilities after softmax. 

In our scenario, we consider terms to be words, entities, or headings in a table. We also employ negative sampling, to make the training of our models computationally more efficient.

\vspace*{-0.5\baselineskip}
\subsection{Four Variants}
\label{sub:nwe:wer}

%Tables are highly structured. 
%Table elements in a web table include:
%%
%\begin{enumerate}
%    \item \emph{pgTitle}, the title of the web page embeds the table. 
%    \item \emph{secondTitle}, section title of the page that includes the table.
%    \item \emph{caption}, title of the table, which gives brief description of content within the table cells. 
%    \item \emph{colHeadings}, a list column heading labels. 
%    \item \emph{tableCells}, content of the table cells, include the heading row.
%%    \item \emph{cat\_all*}, concatenation of all five elements described above. 
%\end{enumerate}
% 
%Content from the same table or the same table elements is most likely related and shares some similar semantic information. 

We train four different table embeddings, using different table elements as input; these are summarized in Table~\ref{tbl:em} and illustrated in Fig.~\ref{fig:table2vec}.
%In terms of the table elements, we select four types of raw representations to represent the table and for training different neural embeddings. Table~\ref{tbl:em} lists the input (second column), which corresponds to the four embeddings (first column). 
All embeddings are trained using the same neural model, but they differ in (i) what constitutes as a term and (ii) which table elements are used for training.

\begin{description}
	\item[Table2VecW] This method takes all the words appearing in a table into consideration. Specifically, it considers the page title, section title, table caption, table headings, and all table cells; see Fig.~\ref{fig:a}.
	\item[Table2VecH] Instead of using single words, we further leverage the table structure and represent tables as sequences of headings. Each heading is treated as a single term, as is shown in the shadowed area in Fig.~\ref{fig:b}.
	\item[Table2VecE] Tables often contain entities, which are semantically more meaningful than words. Thus, we take  sequences of entities as input, by extracting all entities that appear within table cells; see the shadowed area in Fig.~\ref{fig:c}.
	\item[Table2VecE*] Relational tables describe a set of entities as well as their attributes in the columns. These entities are placed in the \emph{core column}. Table2VecE* considers only entities in the core column of the table, as is shown in Fig.~\ref{fig:d}.
\end{description}
\begin{table}[t]
  \centering
  \caption{Table2Vec embeddings.}\label{tbl:em}
	\vspace*{-0.75\baselineskip}  
    \begin{tabular}{ l llc ll }
    \toprule
     Method & Input & Semantic repr. \\
    \midrule
     Table2VecW & all table data & word embeddings   \\
	 Table2VecH & table headings & heading embeddings  \\     
     Table2VecE & all entities & entity embeddings   \\
     Table2VecE* & core column entities & entity embeddings  \\
    \bottomrule
  \end{tabular}
  \vspace*{-0.75\baselineskip}  
\end{table}
%
%\sz{Using the constructed model introduced in Sec.~\ref{sub:nwe:nmt}, we train the Table2Vec embeddings listed in Table 1 and apply the embeddings to subsequent table-related tasks.}

%\subsubsection{Entity based extraction} (i) \emph{Table2VecE}.  (ii) \emph{Table2VecE*} Instead of extracting all the entities from a table, 
\vspace*{-0.75\baselineskip}
\section{Utilizing Table2Vec Embeddings}
\label{sec:app}

In this section, we extend previous table population and retrieval methods by incorporating the Table2Vec embeddings that we introduced in Sect.~\ref{sec:nwe}. 
For all tasks, we keep our focus on relational tables.  It is assumed that entities mentioned in the table are recognized and disambiguated by linking them to entries in a knowledge base~\citep{Bhagavatula:2015:TEL}.
The table population task is considered in two flavors: row population and column population.  We shall refer to the input table $T$ as \emph{seed table}, in which the set of entities from the core column are referred as seed entities $E$, and the set of headings are denoted as seed headings $L$.

\vspace*{-0.5\baselineskip}
\subsection{Row Population}

Row population is a task of returning a list of entities, based on their likelihood of being added to the core column of the seed table $T$ in the next row.  The ranking is established based on the similarity of a candidate entity $e$ to the seed table entities $E$.
% that are ranked in a descending order with the relevance with $E$, and the top ranking one is most likely to be added into the \emph{}  in the next row. 
In this task, we measure entity similarity by two approaches: using a knowledge base and using Table2Vec embeddings. 

%\vspace*{-0.5\baselineskip}
\subsubsection{Baselines}
\label{sub:sub:bkb}
We employ three probabilistic ranking methods from~\cite{Zhang:2017:ESA} as our baselines, which rank candidate entities according to $P(e|E)$.  Candidate entity selection is done as in~\cite{Zhang:2017:ESA}.
\begin{description}
    \item[BL1] Entity similarity is measured based on the similarity of relations of $e$, obtained from RDF triples, and those of the seed tables entities $E$.    
%    A relation of an entity $e$ is given by the remainder of a rdf triple with the $e$ excluded.
    \item[BL2] It uses the Wikipedia Link-based Measure~\citep{Milne:2008:ELM} to estimate the semantic relatedness of entities based on their outgoing links (in the knowledge base).
    \item[BL3] It relies on the Jaccard similarity between outgoing links of entities.
\end{description}

%\vspace*{-0.5\baselineskip}
\subsubsection{Using Table2Vec embeddings}
\label{sub:sub:mte}

Recall that we have two entity embeddings, Table2VecE and Table2VecE*. The former is trained on all entities contained in the table, while the latter considers only entities in the core column.  Given that the row population task focuses on the core column, we employ the Table2VecE* embeddings here.
%Table2VecE trains entities from all table cells, but for this task, entities not from the core column are most likely \sz{intruducing} noise for our training corpus. 
%Thus, Table2VecE* are employed for this task.
We measure the similarity of each candidate entity $e$, against the seed entities $e' \in E$, using the cosine similarity of their respective embedding vectors:
%Note that our candidates are selected by returning top-$k$ ranked entities based on relevance between $e'$ and $e$.
%
\begin{equation}\label{eq:4}
    sim(e, E)= \frac{1}{|E|}\sum_{e' \in E}sim(e, e') = \frac{1}{|E|}\sum_{e' \in E}\frac{\vec {v_{e}} \boldsymbol{\cdot} \vec{v_{e'}}}{\lVert{\vec {v_{e}} \rVert \lVert{\vec {v_{e'}} \rVert}}} ~,
\end{equation}
where $|E|$ is the size of seed entity set, and $\vec {v_{e}}$ and $\vec {v_{e'}}$ are the embedding vectors of the candidate and seed entities, respectively. 

We then combine the baseline similarity with the Table2Vec-based similarity using the following linear mixture: 
\begin{equation}
    P(e|E) = \alpha \, P_{KB}(e|E) + (1 - \alpha) \, P_{emb}(e|E) ~,
    \label{eq:rp:comb}
\end{equation}
where $P_{\mathit{KB}}$ is the similarity measured using the knowledge base and $P_{\mathit{emb}}$ is based on table embeddings, and equals to Eq.~\eqref{eq:4}. 
%\ld{Note that here we employ the candidates in~\ref{sub:sub:bkb}}.

%\vspace*{-0.5\baselineskip}
%
\subsection{Column Population}

Column population is the task of returning a ranked list of headings, $l_{1},\ldots,l_{k}$, given a seed table $T$.  The returned headings are ranked based on their relevance to the seed headings $L$. Similarly to row population, we consider two heading similarity measures.

%\vspace*{-0.5\baselineskip}
\subsubsection{Baseline}
The baseline method, using a table corpus, is taken from~\cite{Zhang:2017:ESA}.  First, relevant tables are retrieved from the table corpus. Then, the probability of a candidate heading being relevant $P(l|L)$ is estimated based on the occurrences of that heading in relevant tables.  %Table relevance is established using 

%\vspace*{-0.5\baselineskip}
\subsubsection{Using Table2Vec embeddings}
We use embeddings trained on table headings, Table2VecH, for heading relevance estimation. Similarly to row population, we measure the cosine similarity between the embedding vectors of the candidate heading $l$ and seed headings $l'\in L$. 
Then, the baseline estimate is combined with the embedding-based similarity using: 
\begin{equation}
    P(l|L) = \alpha \, P_{\mathit{KB}}(l|L) + (1 - \alpha) \, P_{\mathit{emb}}(l|L) ~.
    \label{eq:cp:comb}
\end{equation}
%
%where the computing of $P_{\mathit{emb}}(l|L)$ follows analogously to Eq.~\ref{eq:4}.}

\vspace*{-0.5\baselineskip}
\subsection{Table Retrieval}

Table retrieval is the task of returning a ranked list of tables in response to a keyword query $q$, based on their relevance to $q$. 
For this task, we employ a feature-based method as a baseline, which is referred to as the LTR method in~\citep{Zhang:2018:AHT}. We utilize the word-based and entity-based table embeddings, Table2VecW and Table2VecE, to compute additional semantic matching features.  Specifically, each type of embedding contributes four features, for each of the similarity methods in~\citep{Zhang:2018:AHT}.

Given that both the table and query are vectors now, we compute cosine similarity to measure relevance. For comparison purposes, we employ both methods in~\cite{Zhang:2018:AHT}: \emph{early fusion} and \emph{late fusion}. For the former method, query-table relevance is measured between the centroid of query term vectors and the centroid of table term vectors. The latter method computes pairwise cosine similarity between table terms ($\vec t_{j}$) and query terms ($\vec q_{i}$) first, and then aggregates those results. Here, query-table relevance is measured using an aggregator function, which can be: (i) maximum of $cosine(\vec q_{i}, \vec t_{j})$, (ii) sum of $cosine(\vec q_{i}, \vec t_{j})$ (iii) average of $cosine(\vec q_{i}, \vec t_{j})$. In this paper, we combine all four measures (i.e., early fusion and late fusion using max, sum, and avg aggregators) to yield the final similarity score. For performance comparison, we employ pre-trained Graph2Vec~\cite{Ristoski:2016:RGE} and Word2Vec embeddings~\citep{Mikolov:2013:DRW}.

\section{Evaluation}
\label{sec:eva}

In this section, we formulate our research questions (Sect.~\ref{sec:eva:req}), discuss our
experimental setup (Sect.~\ref{sec:eva:exs}), and then present our results and analysis for the three tasks (Sects.~\ref{sec:eva:row}--\ref{sec:eva:ret}).

\vspace*{-0.25\baselineskip}
\subsection{Research Questions}
\label{sec:eva:req}
We address the following research questions: 
\begin{description}
    \item[RQ1] Can Table2Vec improve table population performance against state-of-the-art baselines?
    \item[RQ2] Does the training of word embeddings specifically on tables, as opposed to news, affect retrieval performance?
    \item[RQ3] Which of the semantic representations (entity vs. word embeddings) performs better in table retrieval?
\end{description}

\begin{table}[t]
    \centering
    %\small
    \caption{Statistics for Table2Vec embeddings. Neg is short for negative sampling (measured in number of words).}
    \vspace*{-0.75\baselineskip}
    \begin{tabular}{l@{~~}r@{~~}rcc}
    \toprule
    \textbf{Embedding} & \textbf{Total terms} & \textbf{Unique terms} & \textbf{Neg} & \textbf{Win\_size}\\
    \midrule
    Table2VecW & 200,157,990 & 1,829,874 & 25 & 5\\
%    \hline
    Table2VecH & 7,962,443 & 339,433 & 25 & 20 \\
%    \hline
    Table2VecE & 24,863,683 & 2,159,467 & 25 & 50 \\
%    \hline
    Table2VecE* & 5,367,837 & 1,285,708 & 25 & 50\\
    \bottomrule
    \end{tabular}
    \label{tbl:emb:statistic}
    \vspace*{-0.75\baselineskip}
\end{table}
%

%
%
%\input{other-inputs/fig-eva-gt.tex}
%

%
%\vspace*{-0.5\baselineskip}
\subsection{Experimental Setup}
\label{sec:eva:exs}
For table population, we use Mean Average Precision (MAP) as the main metric and Mean Reciprocal Rank (MRR) as a supplementary metric for performance evaluation.  Table retrieval performance is evaluated by Normalized Discounted Cumulative Gain (NDCG) with a cut-off at 10 and 20. To test significance, we use a two-tailed paired t-test and write $\circ$ to denote not significant, and $\dag$/$\ddag$ to denote significance at the 0.05 and 0.01 levels, respectively.

We use the Wikipedia Tables corpus~\citep{Zhang:2017:ESA}, which contains 1.6 million high-quality relational tables, both for training the Table2Vec embeddings and for the retrieval experiments. %We detail the raw representations for training in Sect.~\ref{sec:nwe}. Additionally, 
For the word-based embedding, Table2VecW, we filter out empty strings, numbers, HTML tags, and stopwords from the raw text during training to obtain a better representation. For Table2VecH, we employ no normalization for the headings, i.e., ``year(s),'' ``year:,'' and ``year'' will be treated as different headings in our experiment. Table~\ref{tbl:emb:statistic} shows the statistics of different Table2Vec embeddings. DBpedia is used as our knowledge base, which is consistent with the original experiments in~\citep{Zhang:2018:AHT,Zhang:2017:ESA}. The test inputs and ground truth assessments are obtained for the three tasks as follows:

\begin{itemize}
	\item \emph{Row population:} we use the test set from~\citep{Zhang:2017:ESA}. It contains 1000 relational tables, of which each table has at least six rows and four columns. For evaluation, we take entities from the first $i$ rows ($i \in [1..5]$) as seed entities, and the remaining entities as ground truth. The test set contains 21,502 unique entities.
	\item \emph{Column population:} we use the test set from~\citep{Zhang:2017:ESA}, consisting of 1000 relational tables. Headings from the first $j$ columns ($j \in [1..3]$) are taken as seed headings, while the rest constitute the ground truth. There are a total of 7,216 unique column headings. %Figure~\ref{fig:gt:col} illustrates the methodology of column population. 
	\item \emph{Table retrieval:} we use a set of 60 queries (two query subsets, QuerySet 1 and QuerySet 2) and corresponding ground truth relevance labels from~\cite{Zhang:2018:AHT}, a total of 3,120 query-table pairs. 
\end{itemize}

\begin{table*}[t]
  \centering
  %\scriptsize
  \vspace*{-0.75\baselineskip}
  \caption{Row population performance. Statistical significance is tested against the respective baseline.}
  \vspace*{-0.75\baselineskip}
    \begin{tabular}{ l  llc  llc  llc  llc  ll }
    \toprule
    & \multicolumn{14}{c}{\textbf{\#Seed entities ($|E|$)}} \\ 
    \textbf{Method} 
       	& \multicolumn{2}{c}{\textbf{1}} & 
    	& \multicolumn{2}{c}{\textbf{2}} & 
    	& \multicolumn{2}{c}{\textbf{3}} & 
    	& \multicolumn{2}{c}{\textbf{4}} & 
    	& \multicolumn{2}{c}{\textbf{5}} \\
    \cline{2-3} \cline{5-6} \cline{8-9} \cline{11-12} \cline{14-15}
    & \textbf{MAP} & \textbf{MRR} & 
    & \textbf{MAP} & \textbf{MRR} & 
    & \textbf{MAP} & \textbf{MRR} & 
    & \textbf{MAP} & \textbf{MRR} & 
    & \textbf{MAP} & \textbf{MRR} \\
    \midrule

    BL1
        & 0.4360 & 0.5552 &
        & 0.4706 & 0.5846 &
        & 0.4788 & 0.5856 &
        & 0.4786 & 0.5779 &
        & 0.4711 & 0.5618 \\
    BL2
        & 0.2612 & 0.4779 & 
        & 0.2778 & 0.4887 &
        & 0.2845 & 0.4811 &
        & 0.2846 & 0.4808 &
        & 0.2817 & 0.4689  \\
    BL3
        & 0.2912 & 0.5024 &
        & 0.3024 & 0.4927 &
        & 0.3028 & 0.4815 &
        & 0.2987 & 0.4780 &
        & 0.2910 & 0.4609  \\
%    \hline
%    Table2VecE*
%        & 0.4982$^\ddagger$ & 0.7623$^\ddagger$ &
%        & 0.5522$^\ddagger$ & 0.8081$^\ddagger$ &
%        & 0.5598$^\ddagger$ & 0.7993$^\ddagger$ &
%        & 0.5543$^\ddagger$ & 0.7787$^\ddagger$ &
%        & 0.5476$^\ddagger$ & 0.7744$^\ddagger$ \\
    \hline
    BL1 + Table2VecE*
        & \textbf{0.5581$^{\ddagger}$} & 0.7414$^{\ddagger}$ &
        & \textbf{0.6147$^{\ddagger}$} & 0.8141$^{\ddagger}$ &
        & \textbf{0.6400$^{\ddagger}$} & 0.8424$^{\ddagger}$ &
        & \textbf{0.6524$^{\ddagger}$} & 0.8427$^{\ddagger}$ &
        & \textbf{0.6533$^{\ddagger}$} & 0.8372$^{\ddagger}$  \\
    BL2 + Table2VecE*
	    & 0.5461$^{\ddagger}$ & {0.7710$^{\ddagger}$} &
	    & 0.6027$^{\ddagger}$ & \textbf{0.8317$^{\ddagger}$} &
	    & 0.6187$^{\ddagger}$ & 0.8440$^{\ddagger}$ &
	    & 0.6217$^{\ddagger}$ & 0.8389$^{\ddagger}$ &
	    & 0.6223$^{\ddagger}$ & \textbf{0.8410$^{\ddagger}$} \\
	BL3 + Table2VecE*
	    & 0.5487$^{\ddagger}$ & \textbf{0.7728$^{\ddagger}$} &
	    & 0.6049$^{\ddagger}$ & {0.8294$^{\ddagger}$} &
	    & 0.6218$^{\ddagger}$ & \textbf{0.8482$^{\ddagger}$} &
	    & 0.6249$^{\ddagger}$ & \textbf{0.8435$^{\ddagger}$} &
	    & 0.6251$^{\ddagger}$ & 0.8395$^{\ddagger}$ \\
        \bottomrule
  \end{tabular}
  \label{tbl:rows:rank}
  \vspace*{-0.5\baselineskip}
\end{table*}
\begin{table}[t]
  \centering
\footnotesize
  \caption{Column population performance. Statistical significance is tested against the baseline. BL is short for baseline, and TH is short for Table2VecH.}
  \vspace*{-0.75\baselineskip}
    \begin{tabular}{ l  llc  llc  llc  llc  ll }
    \toprule
    & \multicolumn{8}{c}{\textbf{\#Seed column labels ($|L|$)}} \\ 
    \textbf{Method} 
    	& \multicolumn{2}{c}{\textbf{1}} & 
    	& \multicolumn{2}{c}{\textbf{2}} & 
    	& \multicolumn{2}{c}{\textbf{3}} \\
    \cline{2-3} \cline{5-6} \cline{8-9} 
    & \textbf{MAP} & \textbf{MRR} & 
    & \textbf{MAP} & \textbf{MRR} & 
    & \textbf{MAP} & \textbf{MRR} \\ 
    \midrule

	BL
	& 0.2507 & 0.3753 &
	& 0.2845 & 0.4037 &
	& 0.2852 & 0.3552 \\

    BL + TH
    & \textbf{0.2551$^\circ$} & \textbf{0.3796$^\circ$} &
    & \textbf{0.3322$^\ddagger$} & \textbf{0.4400$^\circ$} &
    & \textbf{0.4000$^\ddagger$} & \textbf{0.5080$^\ddagger$} \\
        
    \bottomrule
  \end{tabular}
  \label{tbl:column:rank}
\end{table}

\subsection{Row Population}
\label{sec:eva:row}
%\vspace*{-0.25\baselineskip}
%\subsubsection{Result}
The row population results are listed in Table~\ref{tbl:rows:rank}. The top three lines show the results of the baselines from the literature.  The bottom three lines are the results of combining the baselines with Table2VecE*.  
Note that the combination involves a mixture parameter $\alpha$ (cf. Eq.~\eqref{eq:rp:comb}).  To understand the potential of using table embeddings, we perform a grid search in steps of 0.1 for the value of $\alpha$, and report results using the $\alpha$ value that yielded the best MAP score.  The best performing $\alpha$ values for BL1, BL2, and BL3 are 0.4, 0.0, and 0.1, respectively. This means that the second baseline does not contribute at all to the combination.  %That is, the bottom two rows of Table~\ref{tbl:rows:rank} are based only on Table2VecE*, hence the scores in these two rows are identical.

Overall, we find that the combined methods outperform the respective baselines substantially and significantly ($p < 0.01 $). BL1 + Table2VecE* yields the best performance in terms of MAP.  It is worth pointing out that the performance of this combined method improves more with more seed entities than the baseline BL1, which reaches its peak already after two seed entities.  This indicates the seed entities are better utilized in our embedding-based method.

%Out of the three baselines, BL1 performs far better than the other two in terms of both MAP and MRR. This indicates relations given by RDF triples are more beneficial for capturing entity similarity information. 
%For the combined methods, we only report the best performance at bottom block of Table~\ref{tbl:rows:rank}, \ld{and $\alpha$ is 0.4/0/0 for BL1/BL2/BL3 respectively which means the Table2VecE* is complementary with BL1 but not the other two baselines, and it also indicates Table2VecE* benefits from the candidate selection method by three literature baselines}. As observation, the combinations significantly outperform all four baseline methods ($p < 0.01$) and Table2VecE*. This means Table2VecE* and the candidates selection method of baselines complement each other. Furthermore, (BL1 + Table2VecE*) gives us the best performance in terms of MAP, which is consistent with above that BL1 gives the most relatedness information between entities. We also notice that the performance of combination methods improves with more seed entities, which indicates the seed entities are well utilized.
% without a decrease observed in baselines when the input numbers exceed $3$.

\vspace*{-0.75\baselineskip}
\subsection{Column Population}
\label{sec:eva:col}

Table~\ref{tbl:column:rank} shows column population performance.  We find that the combined method involving Table2VecH significantly outperforms the baseline method ($p < 0.01$) in terms of MAP when $|L|>1$. For $|L|=3$ it achieves substantial and significant improvements ($p < 0.01$) both in terms of MAP and MRR.  Moreover, while the baseline performance does not improve with more seed headings, the combined method can effectively utilize larger input sizes and keeps improving the performance. Combining these findings with the results obtained in Sect.~\ref{sec:eva:row}, we answer RQ1 positively.
%According to Fig.~\ref{fig:col:alpha}, 
The interpolation parameter (cf. Eq.~\eqref{eq:cp:comb}) that yielded the best performance for the combined method is $\alpha=0.01$, which indicates Table2VecH similarity is assigned much higher importance than the baseline.

%For both methods, performance improves along with more seed column labels, because more information is given for determining the related labels. 
%This phenomenon is consistent with that in our row population task.

%\vspace*{-0.5\baselineskip}
%
%
\begin{table}[t]
  \centering
  %\small  
  \caption{Table retrieval performance. Statistical significance is tested against the baseline.}
  \vspace*{-0.75\baselineskip}
  \begin{tabular}{ l  llll }
    \toprule 
	\textbf{Method} &\textbf{NDCG@10} & \textbf{NDCG@20} \\
	\midrule
	Baseline & 0.5456 & 0.6031 \\
	Baseline + Word2Vec  & 0.6006$^\dagger$  & 0.6588$^\dagger$  \\
	Baseline + Graph2Vec & 0.5764$^\circ$ & 0.6340$^\circ$  \\
	Baseline + Table2VecW  & 0.6096$^\ddagger$ & 0.6505$^\dagger$ \\
	Baseline + Table2VecE & 0.5569$^\circ$ & 0.6161$^\circ$  \\	
		
    \bottomrule
  \end{tabular}
  \label{tbl:results}
  \vspace*{-1\baselineskip}
\end{table}
\subsection{Table Retrieval}
\label{sec:eva:ret}
To answer RQ2 and RQ3, we list the table retrieval results in Table~\ref{tbl:results}. For Graph2Vec and Table2VecE, we achieve improvements over the baseline but these are not statistically significant. Table2VecW and Word2Vec have very comparable performance to each other and they outperform all other methods and significantly improve over the baseline method ($p < 0.01$). The lack of difference between the two indicates that it does not make a difference for the table retrieval task whether word embeddings are trained specifically on tables or not (RQ2).
As for the different semantic representations (RQ3), these results show that word embeddings are more beneficial for table retrieval than entity embeddings. 

\section{Conclusion and future work}
\label{sec:con}

In this paper, we have introduced Table2Vec, a neural language model for training word and entity embeddings on various table elements. These embeddings have been utilized in three particular table-related tasks, and have been shown to significantly improve retrieval effectiveness. We have derived these embeddings particularly from a Wikipedia tables corpus, which contains only high-quality relational tables. In the future, we wish to extend our research to other table corpora, as well as to other types of tables.

\bibliographystyle{ACM-Reference-Format}
\bibliography{00paper}

\end{document}